# The Hidden Fragility of Complex Systems— Consequences of Change, Changing Consequences

James P. Crutchfield
Complexity Sciences Center
Physics Department
University of California at Davis
Davis, California 95616 USA

*Abstract*: Short-term survival and an exuberant plunge into building our future are generating a new kind of unintended consequence—hidden fragility. This is a direct effect of the sophistication and structural complexity of the socio-technical systems humans create. It is inevitable. And so the challenge is, How much can we understand and predict about these systems and about the social dynamics that lead to their construction?

**Truly Complex Systems**:

Recent events cannot help but lead one to question the social environment and technological world we are constructing for ourselves.

After decades of building a new world economic order, the data is in: The Fall 2008 near-collapse of the global financial system and its heart-wrenching impacts are empirical evidence that pure-market ideology does not work as a design principle for the world's economies. Historically, this design principle was justified in terms of the Efficient Market Hypothesis [1]—markets in their collective behavior will find the unique, optimal equilibrium condition that homogeneously maximizes human welfare. Sadly, this view is a theoretical artifact of experimentally ungrounded models. The mismatch between ideology and reality is desperately large.

The same design principles were championed in the corporate reorganization facilitated by the market deregulation movement over the same decades. Masquerading as concern for the shareholder, the lack of constraint resulted in new levels of mismanagement and a scale of market manipulation rare in history. The resulting instabilities led to the bankruptcies of Enron and World-Com; two examples of the "largest" (at the time) corporate collapses.

The lack of constraint did engender much creativity in financial instruments—a genuine efflorescence in the sophistication and level of abstraction operating in the world financial system. Aided and abetted by novel computing technologies, innovative strategies to reduce risk, to take one example, emboldened investment firms to over-reach using unusually large amounts of leverage. When the real markets did not meet the instruments' statistical-independence assumptions, the virtual wealth, on which the firms floated, simply evaporated.

For example, the innovation of portfolio insurance is implicated in "Black Monday"—the 19 October 1987 global stock market crash, which was the largest one-day percentage decline ever experienced, up to that time. A latent internal instability, automated in some markets with program trading, was amplified to global scale by the very mechanisms to "insure" investment portfolios. This hidden vulnerability was finally reified when a down-tick in the Hong Kong stock market grew through coupled international markets, spreading worldwide and causing an unprecedented and still somewhat mysterious crash.

Long Term Capital Management (LTCM), a hedge fund founded on just these kinds of financial innovations, provides yet another example. A very large (~100B$) portfolio, heavily leveraged from 5B$ in assets, collapsed when relatively small variations in Russian markets revealed the lack of liquidity it required. Fearing the consequences of the firm's bankruptcy, LTCM was bailed out by its peers and the US Federal Reserve in 1998.

Now years later, these painful examples—Enron, WorldCom, LTCM, and the stock market on Black Monday—and others are almost entirely erased from our collective memory. Then again, perhaps they are particular cases, not worth remembering. Or do they have something to teach us about systems that we humans willfully construct to serve our needs? Their fragilities were unseen and unanticipated. Could they have been? Or is being blind-sided by our own creations inevitable? And, in any case, how are we to understand when our creations are so fragile?

Chagrined, it's nonetheless hard for us to miss the irony of good intentions. Financial instruments invented to reduce risk led to massive failures. To emphasize, innovations in managing risk produced risk on huge scales. And this observation leads us to our first puzzle. How is it that mechanisms on a small scale, with specific functioning and compelling benefits at that level, become exactly the drivers for catastrophic failure on the larger scale? Isn't this a contradiction? Doesn't a contradiction mean that such failures cannot happen? It may be a logical contradiction, but let's face the data from these seemingly accidental experiments. These constructed systems appear to have a dynamic that reshapes small-scale design into surprising, large-scale pattern. In any case, emergent contradiction appears to be fact. This is something to understand.

The basic architecture of fragility is not, as it happens, particular to financial systems. And this may be a good thing, in that a comparative view gives some hope of understanding what is going on. It turns out that many other large-scale engineered systems exhibit similar vulnerabilities.

The pre-9/11 air transportation system is a prime example. From volumes of regulatory, operating, and maintenance documents to the airplanes themselves and the planet-spanning traffic routes, air transportation relies on many interlocking and coordinated subsystems. As a whole it is stunningly sophisticated, as are its components. That sophistication, though, means there are many levels of vulnerability. The sheer size of the network and passenger load preclude a centralized monitoring system. The planes themselves, highly evolved machines, are physically vulnerable to external attack and to the spontaneous failure of only one of the many power and control systems required for them to operate. And, of course, both the transportation network and the planes are operated by humans who make mistakes. The net result is a system, on the one hand, on which economies have developed a dependence and, on the other, which is highly vulnerable to intentional disruption by individuals (either workers or terrorists) acting locally. Air travel's very success translates into its being a target; its sophistication provides the leverage points for being co-opted. The events of 9/11 brought the air transportation system to its knees, inflicting heavy financial losses throughout the US economy.

The air transportation system in its success and planet-spanning organization has produced yet another kind of collateral fragility. The world health system is now more fragile than ever before and in unanticipated ways due to the rapid spread of epidemics through human-contact—contact that is accelerated by air transportation. The result is personal. Our health has become vulnerable to the arrival of once-distant, potentially virulent diseases; such as caused by the H1N1 virus that is now careering around the planet via intercontinental travel.

Despite collateral fragilities, we continue to craft much of modern life around such large-scale transport systems. Our homes and industries are powered via continent-scale electrical power distribution grids. From dozens of generating stations, spread over large geographic areas, each grid delivers its power synchronized to within tens of milliseconds. The very design requirements of long-range power-distribution and maintaining synchronization lead to a kind of coherence in behavior that leaves the grids vulnerable to large-scale failures. In such an architecture, the very strategies to mitigate localized failure, as we have repeatedly rediscovered, can lead through a cascade of cause-and-effect responses to major power outages. As it grows in size and sophistication, the power grid could become less, not more, stable.

So, are the goals and practice of engineering the source of vulnerabilities? No, hidden fragilities are not only the result of humans pursuing their needs. Nature herself is an analogous set of large-scale, interconnected systems. Natural systems are highly structured, being the long-lived products of competitive survival—the inheritance of evolution. And the interconnections between the systems are highly structured themselves, often for similar evolutionary reasons.

Climate change over the last several decades is an example of an interconnection between natural systems, particularly the atmosphere and oceans, and designed systems, such as transportation, power generation, agriculture, manufacturing, and extractive industries. The expected consequences of human-induced climate change are now a common-place: increasing mean global temperature, increasing local climatic variations, poleward movement of agriculture, drought, decreasing access to potable water, sea-level rise and coastal inundation, are some of the primary effects expected over the next century.

Humans are implicated in the systemic fragility that is climate change. This, however, need not be the case. For example, the recent wide-scale emergence of insect-driven deforestation suggests the appearance of a new stabilizing feedback loop that, independent of whether or not it originated through human activities, could very well become autonomous [2]. That is, such interconnected systems can innovate their own patterns which, in turn, lead to new fragilities. And, the patterns, having emerged and stabilized, can preclude mitigation.

At this point, we have a veritable zoo of systemic failures—failures in functioning, failures in design, and, certainly, failures in understanding. The list could be extended, too easily, to include the perceptual exaggerations induced by communications media, increasing economic dependence on the Internet, and the struggles for control mediated by policy-making institutions. Nonetheless, the examples given span a large enough range of system types that we can begin to see commonalities.

**Understanding Truly Complex Systems**:

The world economy, financial markets, air transportation, pandemic disease spread, climate change, and insect-driven deforestation are examples of truly complex systems: They consist of multiple components, each component active in different domains and structured in its own right, interconnected in ways that lead to emergent collective behaviors and spontaneous architectural re-organization.

This is the world we built. Nation states maintain their survival and enhance their well being by participating in international trade. The degree of participation reflects the strength of coupling their internal economies to the external world—the extent being a choice that balances internal needs and externally derived benefits. International trade is managed, in turn, at the largest scale via governmental negotiation, trade organizations, and global finance firms who rely on interconnected national markets. Internally, each state's financial system consists of a number of players, from national reserve and investment banks to insurance companies, institutional investors, mortgage banks, asset-based banks, and workers. The real-economic component—materially productive industries and services—is supported by land, ocean, and air transportation systems. And this synopsis is only one slice. It says nothing of human culture and politics. The more one attempts to describe the social environment and technological systems we have constructed, the more complex they appear and the larger the mystery of their functioning becomes.

How are we to begin to understand truly complex systems and their hidden fragilities? Unfortunately, when it comes to considering how complex systems should be designed and should function, contemporary science and engineering, in their traditional positive and constructive role, largely miss the emergence of fragility. As one looks around in the (rather far-flung) literature, one finds models and results focused on rather the opposite—on the upsides of large-scale systems. There is a philosophy of boosterism for complicated systems: collaboration eclipses individual effort; the collective is more robust [3], smarter [4], and tolerant [5] than the local;

economies are sustainable only if they grow; planet-scale geo-engineering will mitigate climate change. In short, after all these years and painful examples of failure, bigger is still better.

So what's going on? Why the mismatch between the reality of hidden fragility and the techno-optimism that one sees driving science and engineering? We plunge head-long into our future, building socio-technical systems that are ever more complicated. Then, we are surprised when they fail so spectacularly.

**Analyzing Fragility: Functional Pattern Formation**

My premise is that truly complex systems, especially the socio-technical systems humans now construct, are inherently fragile. More to the point, they become so as a natural and inevitable product of how limited cognitive capacity, both at the individual and social levels, affects deploying technological solutions.

How does this happen? As they evolve, systems become more sophisticated—structurally more complex. That is, structural and behavioral correlation accumulates between components and across time. At face value, there is nothing problematic with this. It is a necessary part of building systems, as one commandeers new components and incorporates them, as they begin to work together. This is a simple, accretive view of the organization of complex systems in terms of a dynamic process through which they are created, either naturally, through human design, or both. The key and subtle step occurs, though, when the structural relationships between the components, specifically their dynamical interaction, leads to a spontaneous architectural reorganization as new levels of pattern emerge. That is, not only are individual components structurally complex and interconnected "horizontally" but, through evolution, they become "vertically" nested [6,7,8]. The new patterns represent an increased level of abstraction in the system and reflect increased correlation (more structure) at a new level of organization [9,10]. This new level can take on its own functioning, stabilizing if reinforced by the system as a whole. I call this process *functional pattern formation* [11].

Functional pattern formation raises a difficulty, though. The naive assumption of a system being composed of "modules"—in particular, that the modules are structurally or dynamically independent—fails. When correlation spontaneously emerges, the original components no longer need be "modules". They interface in new ways within the system and can give rise to new, unanticipated behaviors and functions that cross the system. Moreover, these new functions can themselves become commandeered by other parts of the system. And, then, the entire process starts over again, with new levels of organization being constructed out of the existing ones. The lack of apparent modularity that results is the main challenge to understanding and analyzing truly complex systems.

In short, fragility emerges due to increasing structural correlation than spans system degrees of freedom and system degrees of abstraction. Fragility is hidden from us because it is emergent.

The argument here draws a connection between being "more structured" and being "more fragile". There is a list of technical conditions that must be met to rigorously translate from the former to the latter. Not all interventions or perturbations manifest fragility and lead to failure. Nonetheless, we intuitively appreciate the connection. We experience it all the time: Systems that are highly structured, complex in the sense I mean, are easy to destroy [12]. The long list of examples above reinforces the intuition, but at some point one must get down to brass tacks to test the hypothesis of fragility emerging through functional pattern formation. By what mechanisms does fragility emerge?

**Lessons from Complex Systems**:

What can we learn from the contemporary study of complex systems and its tools, as found in nonlinear dynamics, patten formation theory, statistical mechanics, and the like?

The first general thing to note is that many of the observed behaviors of truly complex systems are perfectly consistent with the behaviors and organizations that general systems can produce. This could not be said three centuries ago. Prior to the modern era of complex systems, the behavior of the example systems above could only be interpreted as due to pure chance or, in effect, as not having a mechanistic explanation. We now know better. Complex systems have particular internal mechanisms that can be modeled and whose workings we understand. An important side result is that we also now know that there are strict limits to what can be predicted.

For example, one of the earliest lessons from complex systems is the existence of deterministic chaos—perfectly deterministic systems that over the long-term generate complicated and random-seeming behavior. It turns out there are a number of important properties of chaotic systems that underlie hidden fragility.

One is the exponential amplification of small effects. In the present case—networked systems with many degrees of freedom and many layers of organization—this sensitivity manifests itself as the rapid propagation of information across the system. Above, this was referred to as failures that cascade across system components.

Yet another consequence is that impending failure has a signature. In particular, as a system reorganizes itself from one stable behavioral regime to another, following the process of functional pattern formation, fluctuations and noise will be amplified. That is, to exchange the stability of one kind of behavior for another, the system must pass through a condition of neutral stability. There, external perturbations filter more easily through the system, whose behaviors will fluctuate more wildly than before or after the transition.

One general consequence of deterministic chaos—unlike the above, a limitation—is that component interconnections and internal nonlinearities typically mean that most emergent properties, including fragility, cannot be predicted in advance [6,11]. In this case, one appeals to model building and simulation to ferret out the emergent properties. Or, one even appeals to new meth-ods to automate this process [13]. To the extant one can represent a system's overt complexity in a model, the better the position from which to understand, forecast, and intervene.

Another, somewhat different insight from complex systems is more recent and addresses the collective behavior of groups of intelligent agents. Curiously, when intelligence is added to agents the group behavior tends to become more complicated and, in some cases, chaotic. Why? At first blush, agent intelligence is good. It allows for increased memory and sophisticated strategies. The result, though, is that, by anticipating each others' moves, agents start to "chase" each other's changing strategies. The group behavior starts to oscillate; when, in contrast, simple-minded agents would or could not adapt dynamically. Technically stated, dynamical systems consisting of adaptive agents typically do not tend to a mutually beneficial global condition—they cannot find the Nash Equilibrium [14]. The lesson is that dynamical instability is inherent to collectives of adaptive agents.

A key step in understanding complex systems is to monitor how structured they are. In the case of functional pattern formation, the amount of structure increases and one needs to be able to measure this. Computational mechanics [6,15] gives a way to define and measure the degree of organization of a complex system by answering three questions: (i) How much historical information does a system store, (ii) In what architecture is that information stored, and (iii) How is the stored information used to produce future behavior? The temporal evolution of these measures is itself a useful diagnostic for truly complex systems, especially those that through evolution and adaptation build up internal structures and increase in fragility.

The truly complex systems described above can be analyzed along these lines. Better than giving a detailed explication here, though, is to listen to an earlier voice. The philosopher Alfred North Whitehead captured the essential character of evolving, adapting systems most elegantly, when in the 1920s he considered the domain of human social organization [16]:

> The social history of mankind exhibits great organizations in their alternating functions of conditions for progress, and of contrivances for stunting humanity. The history of the Mediterranean lands, and of western Europe, is the history of the blessing and the curse of political organizations, of religious organizations, of schemes of thought, of social agencies for large purposes. The moment of dominance, prayed for, worked for, sacrificed for, by generations of the noblest spirits, marks the turning point where the blessing passes into the curse. Some new principle of refreshment is required. The art of progress is to preserve order amid change, and to preserve change amid order. Life refuses to be embalmed alive. The more prolonged the halt in some unrelieved system of order, the greater the crash of the dead society.

Whitehead also saw parallels in the temporal evolution of structural complexity (and its aesthetic appreciation) in the cognitive-social dynamics of fashion [16]:

The same principle is exhibited by the tedium arising from the unrelieved dominance of fashion in art. Europe, having covered itself with treasures of Gothic architecture, entered upon generations of satiation. These jaded epochs seem to have lost all sense of that particular form of loveliness. It seems as though the last delicacies of feeling require some element of novelty to relieve their massive inheritance from bygone system. Order is not sufficient. What is required, is something much more complex. It is order entering upon novelty; so that the massiveness of order does not degenerate into mere repetition; and so that the novelty is always reflected upon a background of system.

There is a cyclic signature: building up organization that—stabilized, becomes substrate for further innovation, but that, aging, converts to constraint against adaptation—eventually leads to collapse from its very own benefits. It parallels the failure dynamic exhibited by our truly complex systems. One cannot also help but draw parallels with the structural adaption of renewal seen in Nature via ecological succession [17] and hoped for in the economy through the creative destruction of the "business cycle".

**Looking Forward**:

Faced with the increasingly complex socio-technical systems that we collectively build, it is perhaps not hopeful to conclude that fragility is endemic and, worse, that the process which creates it, hides it. Nonetheless, I am hopeful, since the concepts and techniques of complex systems are up to the task of understanding and analyzing hidden fragility.

Hopeful, that is, but for one thing. There is a nagging concern that the public sphere is unable to support the sophisticated discourse required to democratically stop the ignorant creation of conditions for fragility. In other words, the public sphere, itself a truly complex system, does not have sufficient structural complexity. Baldly stated, society has been presented with a problem, but does it support the information processing necessary to compute a solution? I have my doubts. And so, I am optimistic due the available mathematical tools and future likely technical advances. I am less sanguine about how that understanding will be translated into action through political and public processes.

Fortunately, there are hints of increasing self-awareness on the part of the players in some of the domains discussed here. For example, Nobel Laureate Paul Krugman's re-evaluation of the goals and foundations of economics are heartening [18]. Also in the financial sphere, fragility is beginning to receive attention [19]. In the climate domain, there is increasing appreciation of complex-systems subtleties in the recent attention paid to sudden climate shifts [20]: Not all change is gradual and proportionate.

These glimmers are just the beginning. Especially so, if we are to understand the apparent commonality across our truly complex systems and, practically, to stimulate domain experts to talk across their interconnected domains.

A more substantial push is needed for a general theory of complex systems. Of late, in response to criticisms of too much abstraction, the field has taken a respite from this difficult task to focus on disseminating its tools through applications. Now it is time to turn back to the search for general principles. The challenges will not be met by only studying particular cases; conceptual innovation is required. This is particularly necessary, since many of the problems confronting us require action—we will have to intervene. And this brings in a whole new level of understanding how to control complex systems. Of similar importance is understanding time-dependent control or, what above was called, "adaptation". This topic sorely needs attention.

We will also require new tools to manage the vast amounts of data that we can (and should) be collecting from truly complex systems. For example, we need new tools that automate building models, analyzing hierarchical organization, and monitoring structural complexity and fragility [6,13]. Indeed, what is the role of experiment in building our complex socio-technical systems? To date, little or none. But we are clearly engaged, if accidentally, in experimentation. So, at least, we could take better data. This would provide some solace, perhaps only to a vanishingly small degree, for the suffering that follows when our systems fail.

James P. Crutchfield is Director of the Complexity Sciences Center and a physics professor at the University of California at Davis; http://csc.ucdavis.edu/~chaos/. This essay is based on a talk "Terrorizing Complex Systems" presented at the Santa Fe Institute to the Business Network Topical Meeting on "Modeling Terrorism as a Complex Adaptive System", held in Santa Fe, New Mexico, 10 April 2003. It has been updated to include illustrative events since that time. (The talk PDF is available at http://csc.ucdavis.edu/~chaos/chaos/talks.htm.) For the updates, the author acknowledges comments from the participants of SFI's "Forum on Risk", University Club of New York, New York, 18 October 2007. This work was supported in part by the National Academies Keck Futures Intitiative.